\documentclass[twocolumn,aps,prl,showpacs,preprintnumbers]{revtex4}%
\usepackage{mathrsfs}
\usepackage{amsmath}
\usepackage{amsfonts}
\usepackage{amssymb}
\usepackage{amsthm}
\usepackage{graphicx}
\usepackage{color}
\usepackage{graphics}
\usepackage{epsfig}
\usepackage{cancel}
\providecommand{\U}[1]{\protect\rule{.1in}{.1in}}
\usepackage{amsmath}
\usepackage{ulem}
\usepackage{amsfonts}
\usepackage{amssymb}
\usepackage{amsthm}
\usepackage{graphicx}
\usepackage{epsfig}
\usepackage{color}
\usepackage{subfigure}
\usepackage{graphics}

\newcommand{\bfr}{{\bf r}}
\begin{document}
\title{A Band of Critical States in Anderson Localization at Strong
Magnetic Field with Random Spin-Orbit Scattering}
\author{C. Wang$^{1,2}$}
\author{Ying Su$^{1,2}$}
\author{Y.  Avishai$^{3}$}
\email[corresponding author: ]{yshai@bgumail.bgu.ac.il}
\author{Yigal Meir$^{3}$}
\author{X. R. Wang$^{1,2}$}
\email[corresponding author: ]{phxwan@ust.hk}
\affiliation{$^{1}$Physics Department, The Hong Kong University of
Science and Technology, Clear Water Bay, Kowloon, Hong Kong}
\affiliation{$^{2}$HKUST Shenzhen Research Institute, Shenzhen 518057,
China}
\affiliation{$^{3}$Department of Physics, Ben-Gurion University of
the Negev Beer-Sheva, Israel}
\date{\today}
\begin{abstract}
Anderson localization problem for non-interacting two-dimensional
electron gas subject to strong magnetic field, disordered potential
and spin-orbit coupling is studied numerically on a square lattice.
The nature of the corresponding localization-delocalization
transition and the properties of the pertinent extended states depend
on the nature of the spin-orbit coupling (uniform or fully random).
For uniform spin-orbit coupling (such as Rashba coupling),
there is a band of extended states in the center of a Landau
band as in a ``standard" Anderson metal-insulator transition.
However, for {\it fully random} spin-orbit coupling, the familiar
pattern of Landau bands disappears. Instead, there is {\it a central
band of critical states with definite fractal structure} separated
at two critical energies from two side bands of localized states.
Moreover, finite size scaling analysis suggests that for this novel
transition, on the localized side of a critical energy $E_c$, the
localization length diverges as
$\xi(E) \propto\exp(\alpha/\sqrt{|E-E_c|})$, a behavior
which, along with the band of critical states,
is reminiscent of a Berezinskii-Kosterlitz-Thouless transition.
\end{abstract}

\maketitle
Traditionally, non-interacting disordered electronic systems subject
to disordered potential are classified according to  the symmetries of
their Hamiltonian with respect to time reversal (TR) and spin rotation
(SR) transformations. Considering the Hamiltonian as a random matrix
\cite{Wigner,Dyson,Mehta}, its symmetries determine to which random
matrix Gaussian ensemble (also referred to as universality class) it
belongs, orthogonal (both TR and SR symmetries are satisfied),
symplectic (only TR symmetry) or unitary.

This classification is intimately related to one of the most fundamental
concepts in the physics of disordered electronic systems: the Anderson
localization transition (ALT) \cite{review,Huckestein} that is a {\it
quantum phase transition} between localized and extended states in a
disordered system.
The critical dimension for existence or non-existence of ALT is $d=2$.
For $d<2$ there is no ALT while for $d>2$ there is always ALT.
Hence, for two-dimensional electron gas (2DEG), ALT (if it exists) is of
special interest. The scaling theory of localization \cite{Abrahams}
(developed before the discovery of the quantum Hall effect) together
with calculations based on nonlinear sigma model \cite{Friedan,Hikami},
established that for $d=2$, ALT does not exist for the orthogonal and
unitary classes (zero or finite magnetic field, respectively) and does
exist for the symplectic class (finite spin-orbit scattering and zero
magnetic field). After the discovery of the integer Quantum Hall
effect (IQHE) \cite{QHE}, topology was also recognized as a property
determining the pertinent universality class \cite{TKNN}.
Mathematically, the role of topology in the IQHE is quantified by
the occurrence of a topological term in the action of the
corresponding non-linear sigma model\cite{Pruisken}.
In the presence of the topological term it is established that {\it if
SR invariance is respected}, the system is in the IQHE universality
class characterized by a Hall transition between localized and critical
states occurring at discrete energies. What is less clear (and motivates
the present study) is what happens in the presence of the topological
term {\it when SR invariance is broken}. To the best of our knowledge,
there is so far no rigorous extension of the non-linear sigma model for
an IQHE system in the presence of strong spin-orbit coupling (SOC).

In this work we investigate (numerically) the nature of transition
between localized and extended states for a disordered 2DEG subject to
strong (perpendicular) magnetic field in which SR invariance is broken
due to SOC. In Ref.~\cite{AM}, the nature of states between Zeeman split
states of the first and second Landau levels was investigate for weak
random SOC, and a percolation like ALT transition has been established.
Here we neglect the Zeeman energy and show that the nature of the SOC
(uniform or fully random) dramatically affects the pertinent transition.
Our main results  are:
1) For {\it uniform} SOC (such as Rashba coupling due to a uniform
electric field), the pattern of separated Landau bands persists.
Focusing attention on the lowest Landau band centered at an energy
$E_0$, it is shown that there are two energies $E_{c_1}<E_0< E_{c_2}$
that are critical in the sense that
the states $\psi_E(\bfr)$ for $E_{c_1}<E<E_{c_2}$ are {\it metallic},
while for $E \notin [E_{c_1},E_{c_2}]$, they are localized.
This is a ``usual" ALT between a band of
localized states and a band of metallic states.
Finite size scaling analysis indicates that the critical exponent for
the divergence of the localization length is close to that of the IQHE.
2) However, for {\it fully random} spin-orbit coupling, the structure of
broadened Landau bands that is the hallmark of the IQHE is completely
washed out. In turn, there is a broad {\it band of critical states} with
definite fractal structure (the fractal dimension equals $1.82\pm0.02$).
This band is separated in two critical energies
$E_{c_1} < E_{c_2}$ from two narrow side bands of localized states.
Finite size scaling analysis suggests that for this novel transition,
the localization length at energy $E$ (on the localized side) diverges as
$\xi(E) \propto\exp(\alpha/\sqrt{|E-E_c|})$, a behavior
reminiscent of a Berezinskii-Kosterlitz-Thouless (BKT) transition \cite{BKT}.

To substantiate our claims we consider a tight-binding Hamiltonian for
2DEG on a square lattice of length $L$ and width $M$ (the lattice
constant is set to unity),
\begin{align}
\begin{split}
H=\sum_{i,\sigma}\epsilon_{i}c_{i,\sigma}^{\dagger}c_{i,\sigma}+
\sum_{<ij>,\sigma,\sigma'} \exp(i\phi_{ij})V_{ij}(\sigma,\sigma')
c_{i,\sigma'}^{\dagger}c_{j,\sigma}.
\end{split}
\label{Hamiltonian}
\end{align}
Here $i=(n_i,m_i)$ is a lattice site specified by integer coordinates $n_i$ and
$m_i$ with $1 \le n_i \le L$ and $1 \le m_i \le M$.
$c_{i,\sigma}^\dagger$ ($c_{i,\sigma}$) is electron creation
(annihilation) operator of spin $\sigma=\pm$ on site $i$.
The on-site energy $\epsilon_{i}$ are random and uniformly
distributed in the range of $[-W/2,W/2]$, so that $W$ measures the
degree of randomness. The symbol $<ij>$ indicates that $i$ and $j$ are
 nearest neighbor sites. The magnetic field is introduced
through the Peierls' substitution \cite{Peirls} by endowing  the
hopping coefficient with a phase,
$\phi_{ij}=\tfrac{e}{\hbar} \int_{i}^{j}\vec{A}\cdot d\vec{l}$, where
$\vec{A}$ is the vector potential. For a constant magnetic field $B$
the sum of phases along four links around a square (the same  for all
squares) is written as $2 \pi \phi $, where $\phi$ is the magnetic
flux  per square in units of the quantum flux $\Phi_0=ch/e$.
Henceforth, the magnetic field $B$ is expressed in terms of $\phi$.

The SOC is encoded by $2 \times 2$ matrices $V_{ij}$ acting in spin space.
We will explore both the case of  constant SOC matrices along the axes
and the case of  fully random SOC matrices. In the case of constant SOC
matrices, they are parametrized as $V_{ij}=V_x$ ($V_y$) for $<ij>$ along
the $x-$direction ($y-$direction). In order to get non-trivial results
due to SOC, one requires $[V_x,V_y] \ne 0$.

The Rashba form of uniform SOC reads,
\begin{equation}
V_x=
\begin{bmatrix}
1 & a \\
-a & 1
\end{bmatrix}
\quad\text{and}\quad
V_y=
\begin{bmatrix}
1 & -ia \\
-ia & 1
\end{bmatrix},
\label{RF}
\end{equation}
where $a$ is a real constant encoding the strength of the SOC
for the Rashba model.
Random SOC is encoded by matrices
$V_{ij} \in $ SU(2) thereby defining the SU(2) model\cite{SU2},
\begin{equation}
V_{ij}=\begin{bmatrix} e^{-i\alpha_{ij}}\cos(\beta_{ij}/2) &
e^{-i\gamma_{ij}}\sin(\beta_{ij}/2)\\ -e^{i\gamma_{ij}}\sin(\beta_{ij}/2)
&e^{i\alpha_{ij}}\cos(\beta_{ij}/2) \end{bmatrix},
\label{RSU2}
\end{equation}
where $\alpha_{ij}$, $\beta_{ij}$ and $\gamma_{ij}$ are random angles.
In the full SU(2) model studied here, $\alpha_{ij}$ and $\gamma_{ij}$ are
uniformly distributed in $[0,2\pi]$ and $\sin(\beta_{ij})$ is uniformly
distributed in $[0,1]$.\par

Localization-delocalization transition in an electronic system at zero
temperature is characterized  not only by divergence of the localization
length but also by the nature of the wave functions $\psi_{E_c}(\bfr) $
at the critical energies. These two criteria are independent of each
other, and their analysis usually require two different numerical procedures. \\
\underline{Study of the localization properties:}
We consider
a scattering problem for an electron at Fermi energy $E$ living on a
square lattice of length $L \to \infty$ (along $x$) and finite
width $M$ (along $y$). Periodic boundary conditions are imposed
along $y$ to avoid edge states contribution.

Since the system is quasi one-dimensional, it has a finite localization
length $\lambda_M(E)$ depending on the energy and the system's width $M$.
Using the transfer matrix method we calculate  $\lambda_M(E)$
by a standard iteration algorithm \cite{review,XCXie}.
In our calculations $L>10^6 \gg \lambda_{M}(E)$ and self-averaging
requires relatively small data ensembles to achieve good statistics.
The width $M$ takes values between 32 and 96.
Denoting the normalized localization length by $\bar{\lambda}_M(E)\equiv
\lambda_{M}(E)/M$, the identification of a transition point $E_c$
on the energy axis is guided by the following observations:
1) For metallic (insulating) scattering states,
$ \bar{\lambda}_M(E)$ is an increasing (decreasing) function of $M$.
2) For critical states, $ \bar{\lambda}_M(E)$ is independent of $M$.
3) For energy $E$ close to a critical point $E_c$, $\bar{\lambda}_M(E)$
obeys  a single parameter finite size scaling.
Explicitly, let us denote by $\xi(E) \equiv \lambda_\infty(E)$ the
localization length for a system of width $M \to \infty$.
At the critical energy we expect $\xi(E_c)=\infty$,
and finite size scaling implies $\bar{\lambda}_{M}(E)=f[\tfrac{M}{\xi(E)}
]$ where $f(x)$ is a universal (disorder independent) scaling function.
\par
\begin{figure}
\begin{center}
\includegraphics[width=8.5cm]{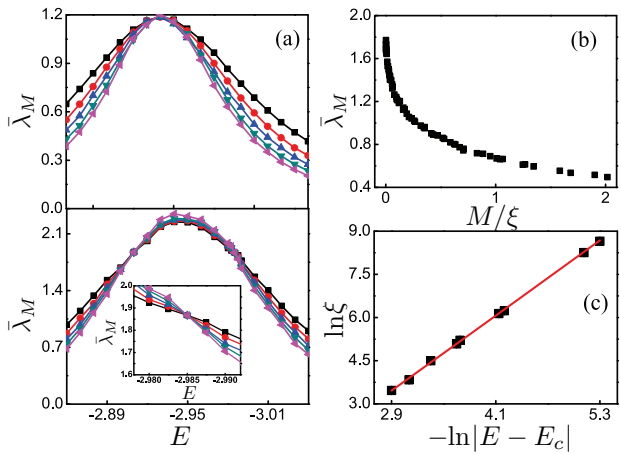}
\end{center}
\vspace{-0.2in}
\caption{(Color online) $\bar{\lambda}_M(E)$ (averaged over
40 disorder realizations), {\it v.s} $E$ for $B=1/5$ and $W=1$.
(a) Without SOC (upper panel) and with constant Rashba SOC,
Eq.~(\ref{RF}) with $a=0.1$, (bottom panel). The system widths (from
top to bottom) are $M=32$ (square), 48 (circle), 64 (up-triangle), 80
(down-triangle), and 96 (left-triangle). The inset of the bottom panel
is a zoom in on the crossing region around $E_c=-2.985$.
(b) When the points $\bar {\lambda}_M(E)$ shown in (a) are expressed
in terms of $x=M/\xi(E)$ (where the localization length $\xi(E)$ is
derived through the numerical procedure), they fall on a smooth curve
thereby display the scaling function.
(c) $\ln \xi(E)$ {\it v.s} $\ln(|E-E_c|)$, $E_c=-2.985$,
for the constant Rashba SOC.
The solid line is linear fit with slope (critical exponent) $\nu=2.2$.
  }
\label{Normal MIT}
\vspace{-0.2in}
\end{figure}

\noindent
\underline{The constant SOC case, Eq.~(\ref{RF}):}
In this case, the pattern of separate LBs remains intact, and our attention
is focused on the lowest LB.
The upper panel of
Fig.~\ref{Normal MIT}(a) displays the quantity $\ln \bar{\lambda}_M(E)$
{\it v.s} $E$ for $B=1/5$ and $W=1$ in the {\it absence} of the SOC, that
is,  $a=0$. It is evident that all curves for different widths $M$
coalesce at the peak energy $E_c$.
On both sides of $E_c$, $\ln \bar{\lambda}_{M}(E)$ {\it decreases}
with $M$, indicating that all states away from $E_c$ are  localized.
However, at the critical point $E_c$, $ \bar{\lambda}_{M}(E)$ is
independent of $M$, and the corresponding states are critical \cite{cwang}.
The bottom panel displays $\ln \bar{\lambda}_M(E)$ for the same values
of $B$ and $W$ as in the upper panel when the constant (Rashba) SOC
strength is $a=0.1$. Here, in contradistinction from the case of zero
SOC, curves of different $M$ cross at {\it two energies}
($E_{c,1}=-2.91\pm 0.01,E_{c,2}=-2.99\pm 0.01$).
Moreover, states between these two  points are {\it metallic}
because $\ln \bar{\lambda}_{M}(E)$ {\it increases} with $M$, indicating
that these are metallic extended states (the corresponding
wave-functions have a trivial fractal structure as for plane-waves).

Now we use the hypothesis of single parameter finite size scaling to
substantiate the criticality of the transition, namely, that for $E$
close to $E_{c,i}$ $\bar{\lambda}_M(E) = f[\tfrac{M}{\xi(E)}]$.
The results are summarized in Fig.~\ref{Normal MIT}(b,c).
As shown in Fig.~\ref{Normal MIT}(b), in the vicinity of the
crossing points, all data points $\bar{\lambda}_M(E)$ for the
Rashba SOC case collapse onto a single smooth scaling curve $f(x)$.
Like in the standard ALT or Hall transitions, $\xi(E)$ diverges as a power,
$|E-E_c|^{-\nu}$, with $\nu$ the localization length critical exponent.
This is substantiated in Fig.~\ref{Normal MIT}(c) that displays
$\ln\xi(E)$ {\it v.s} $\ln|E-E_{c,2}|$ for the Rashba SOC case, Eq.~(\ref{RF}).
The fit to a straight line is rather satisfactory, yielding a slope
$\nu_1=2.2\pm0.1$, that is somewhat smaller than both critical exponents
of the 2D disordered systems $\nu\simeq 2.75$ for the 2D symplectic
symmetry class \cite{SU2} and $\nu\simeq 2.34$ for the IQHE \cite{Huckestein}.
\par

\begin{figure}
\begin{center}
\includegraphics[width=8.5cm]{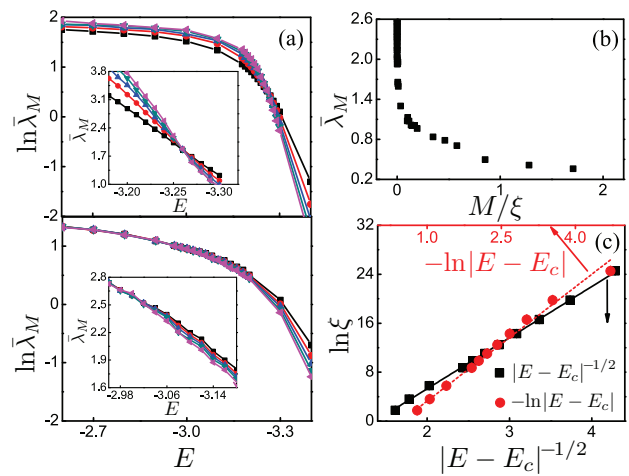}\\
\end{center}
\vspace{-0.2in}
\caption{(color online) (a) $\ln(\bar{\lambda}_M)$ {\it v.s} $E$ of the
SU(2) model with $W=1$ and $M=32$ (square); 48 (circle); 64 (up-triangle);
80 (down-triangle); 96 (left-triangle). The top panel is for $B=0$, and
the inset is a zoom in on the crossing region in linear scale.
The bottom panel is for  $B=1/5$, and the inset is a zoom in on the merging region.
(b) The scaling function obtained from the bottom panel in (a) by collapsing
data of $\bar{\lambda}_M$ near the merging point into a single curve.
(c) $\ln\xi$ {\it v.s} $|E-E_c|^{-1/2}$ for $E_c=-3.001$ (squares).
The solid line is a linear fit with slope $\alpha=8.3\pm 0.3$.
For a comparison, $\ln\xi$ {\it v.s} $-\ln|E-E_c|$ with $E_c=-3.051$
(circles) is also plotted. Larger deviation in the linear fit
(dashed line with goodness-of-fit of $5.2\times 10^{-7}$) indicates that
an interpretation in terms of BKT-type transition (with goodness-of-fit of $8.\times 10^{-3}$) explains the
data better. }
\vspace{-0.3in}
\label{SU2}
\end{figure}

\noindent
\underline{Localization for the fully random SU(2) model, Eq.~(\ref{RSU2}):}
In the absence of a magnetic field, the SU(2) model supports a
standard ALT as shown in the upper panel of Fig.~\ref{SU2}(a) which
plots $\ln(\bar{\lambda}_M(E))$ {\it v.s} $E$ for $W=1$ and various $M$.
All curves cross at $E_c=-3.259$, showing all states of $E\in [-3.259,
3.259]$ are extended because $\ln(\bar{\lambda}_M(E))$ increases with
$M$ as shown clearly in the inset of the enlarged crossing region.
Finite size scaling yields the value
$\nu=2.73\pm0.02$ commensurate with earlier calculations \cite{SU2}.\\
We come now to the main result of the present work.
Switching a strong magnetic field $B=1/5$ one would expect
a  pattern of LB modified due to the presence of SOC. However, what we find
is that the curves $\bar{\lambda}_M(E)$ do not display separate LB peaks,
but, rather, a single band. More remarkably, in contradistinction with
the symplectic case (SOC and $B=0$), the curves $\bar{\lambda}_M(E)$ that
display a localized region for energies $\{ E \}$ near the band edge
(that is, $\bar{\lambda}_M(E)$ decreases with $M$),
{\it do not cross but  merge} as the energy approaches the band center.
This is evident by looking at the bottom panel of Fig.~\ref{SU2}(a) that
displays $\ln(\bar{\lambda}_M(E))$,
averaged over 40 ensembles, as a function of $E$ for $B=1/5$, $W=1$
and various system widths $M$. The inset is a zoom in of the merging region.
For $E< E_c=-3.001$ the system behaves as an insulator where
$\bar{\lambda}_M(E)$ decreases with $M$. But for $E \ge E_c$ all curves
merge, forming a band of critical states for which $\bar{\lambda}_M(E)$
is independent of $M$, and, as we  shall see below, the corresponding
wave-functions cover only a fractal part of 2D space.
This band of critical states  prevails for all energies
$|E|<|E_c|=3.001$, namely the pattern of separate Landau bands
is completely washed out.

To explore the nature of this localization-delocalization transition
we inspect the behavior of $\xi(E)$ on the insulating side $E < E_c$.
Fig.~\ref{SU2}(b) depicts the collapse of all curves $\bar{\lambda}_M(E)$
for different widths $M$, supporting the quantum phase transition
interpretation. However, if one analyzes the divergence of $\xi$ in terms
of a power law, as shown by the red dotted line in the log-log plot in
Fig.~\ref{SU2}(c), the goodness-of-fit \cite{method} to the numerical
data (red circles) is about $5.2\times 10^{-7}$,
which is four orders of magnitude smaller than an acceptable value.
Moreover, the resulting exponent turns out to be $\nu\simeq6.4$, much
larger than any known critical exponent in these kind of systems.
The occurrence of a line of critical points in our
2D system is reminiscent of a critical behavior of the BKT
phase transition \cite{BKT}, $\xi(E)\propto\exp(\alpha/\sqrt{|E_c-E|})$.
Indeed, the fit to a straight line in Fig.~\ref{SU2}(c), which displays
$\ln\xi$ against $1/\sqrt{E-E_c}$, supports the assertion of BKT-type
transition \cite{XCXie}. Quantitavively, the corresponding goodness-of-fit
to the numerical data (black squares) is about $8.\times 10^{-3}$,
far better than the power-law fit, thereby supporting our claim.

\begin{figure}
\begin{center}
\includegraphics[width=8.5cm]{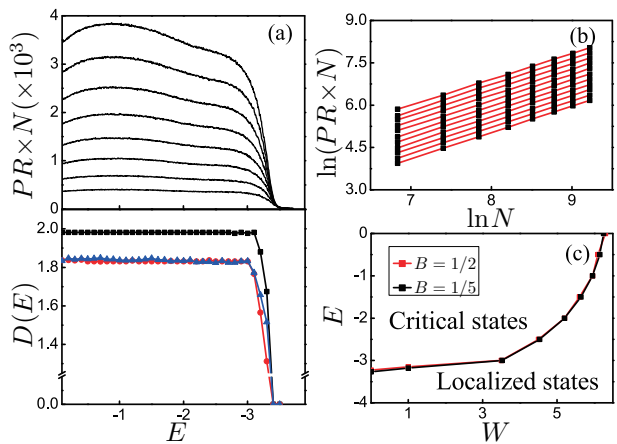}\\
\end{center}
\caption{(color online) (a) Upper panel: $PR(E)\times N$, averaged over
100 samples, as a function of $E$ of the SU(2) model, Eq.~(\ref{RSU2}),
for $B=1/5$ and $W=1$. The lattice size (from down up) is $M=$30; 40;
50; 60; 70; 80; 90; 100. Lower panel: $D(E)$ {\it v.s} $E$ for $W=1$ and
$B=0$ (black square); $B=1/5$ (red circle); $B=1/2$ (blue triangle).
(b) The quantity $\ln(PR(E)\times N)$ is plotted {\it v.s} $\ln N$ for
$E=-2.9,-2.8,\cdots,-2$.
The solid lines are the linear fits of the data with slopes $0.91\pm0.01$.
(c) Phase diagram of the SU(2) model for $B=1/2$ (red circle) and $B=1/5$
(black square) in the $E-W$ plane. Above $W_c=6.3$, all states are
localized.  }
\label{D(E)}
\vspace{-0.2in}
\end{figure}
\noindent
\underline{Fractal structure of critical states:}
To confirm the criticality of all states along the merging line, we
compute, by exact diagonalization, the normalized electron wave
functions $\psi_E(i)$
on a square lattice of {\it finite extent}
(for convenience we take a rectangle of area $M(M+1)$).
Here $i=(n_i,m_i)=1,2,\ldots,N=M(M+1)$ is a site on this lattice).
According to our previous discussion, the wave functions for energies
$E>-3.001$ belong to the band of critical states. Their fractal nature
can be confirmed by computing the participation ratio,
\begin{align}
\begin{split}
PR(E)=\frac{1}{N\sum_{i}|\psi_{E}(i)|^{4}}.
\end{split}
\label{PR}
\end{align}
For a state whose wave-function occupies a fractal space of
dimension $D(E)$ \cite{PR}, $PR$ scales with $N$ as $PR\propto N^{-1+D(E)
/2}$, with $D(E) \ge 0$. For a localized state, $PR\propto N^{-1}$.
The upper panel of Fig.~\ref{D(E)}(a) displays $PR\times N$ as a function of
energy for $B=1/5$, $W=1$ (same as those in Fig.~\ref{SU2}) and various $M$.
All curves merge for $E<E_c$ (localized region), showing the independence
of $PR\times N$ on $M$. Thus those wavefunctions of energy less than $E_c$
are indeed localized. For $0\ge E>E_c$, $PR\times N$ increases with $M$.
To demonstrate that these states are critical with a non-trivial fractal
structure, Fig.~\ref{D(E)}(b) displays $\ln [PR]$ against  $\ln N$ for 10
different energies $E=-2.9;-2.8;-2.7;\ldots; -2$. They are almost parallel
to each other with a common slope of $0.91\pm0.01$, indicating fractal
wave-functions of fractal dimension $D(E)=1.82\pm0.02$ for those states.

The fractal nature of the critical states is universal in the sense that
$D(E)$ does not depend neither on the magnetic field (as long as it is
strong enough) nor on energy (as long as $E>E_c$). This is substantiated in
the bottom panel of Fig.~\ref{D(E)} (a) where $D(E)$ is plotted {\it v.s}
$E$ for $W=1$ and $B=1/5$ (red circles) and $B=1/2$ (blue triangles).
It is instructive to compare the fractal properties of the critical
wave-functions discussed above with those of the wave-functions for
the  SU(2) model {\it at zero magnetic field} (that is, the metallic side
for the symplectic ensemble).
The fractal dimension of these wave-function as a function of $E$ is
also shown in the bottom panel of Fig.~\ref{D(E)}(a) (black squares).
In contrast with the case $B=1/5$
for which the fractal dimension is shown to be $1.82\pm0.02$,
the extended states in the absence of the
field are metallic ($D(E)=2$) and occupy the entire lattice.
Our fractal dimension is somewhat bigger than $D=1.75$ for the critical
state of IQHE systems \cite{Mirlin} and $D=1.66\pm 0.05$ for
the citical state of SU(2) model at zero field \cite{Obuse}.

The critical point $E_c$ that marks the edge of the band of critical
states clearly depends on the strength of disorder $W$. The larger
is $W$, the smaller is $E_c$. On the other hand, for strong enough
field, $E_c$ is virtually independent on the magnetic field.
It is then useful to draw a phase diagram of the SU(2) model where
the line $E_c(W)$ separates regions of localized and critical states.
This analysis is carried out and the result is presented in
Fig.~\ref{D(E)}(c) where $E_c(W)$ is plotted {\it v.s}
$W$ for $B=1/5$ (black squares) and $B=1/2$ (red circles).
The fact that for $W \ge 6.5$ all states are localized
(albeit in the absence of SOC),
has already been substantiated\cite{cwang}.

In conclusion, the nature of ALT for 2DEGs with potential disorder and
SOC subject to a strong perpendicular magnetic field depends on whether
the SOCs is realized by constant or fully random SU(2) matrices operating
in spin space. For constant  SOC, (such as in the Rashba term induced
by a uniform electric field, Eq.~(\ref{RF})),  there is a normal ALT
separating localized and extended states that form a band of finite width.
The corresponding critical exponent is  similar to that
obtained in the absence of magnetic field for the symplectic ensemble.
On the other hand,  for the fully random SU(2) model of the SOC, 
Eq.~(\ref{RSU2}), the pattern of separated LBs is smeared and the system undergoes
a BKT-type transition separating localized states from critical states.
The localization length diverges as an exponential of an inverse square
root and the critical states form a band of finite width and occupy a
fractal space whose dimension is about 1.82. This is in contrast with zero
field case, where in the presence of fully random spin-orbit scattering
the system undergoes a regular ALT separating localized states from
extended (metallic) states.

\noindent
This work is supported by NSFC of China grant (11374249) and Hong
Kong RGC grants (605413). YA would like to thank the Physics
Department of HKUST for hospitality during 2012-2014.
The research of YA is partially supported by Israeli Science
Foundation Grants 1173/2008 and 400/2012.

\end{document}